\documentclass[11pt]{article}

\usepackage{bm}
\usepackage{amsmath}
\usepackage{amssymb}
\usepackage{fullpage}
\usepackage{graphicx}
\usepackage{bm}
\usepackage{hyperref}
\usepackage{cite}
\usepackage{authblk}
\usepackage{color}

\newcommand{\blue}{\textcolor{blue}}

\PassOptionsToPackage{hyphens}{url}\usepackage{hyperref}

\title{\textbf{The Evolution of Altruistic Rationality Provides a Solution to Social Dilemmas via Rational Reciprocity}}
\date{\today}

\author[1,2,3]{Mohammad Salahshour\thanks{msalahshour@ab.mpg.de}}
\author[1,2,3]{Iain D. Couzin}
\affil[1]{Department of Collective Behaviour, Max Planck Institute of Animal Behavior,
	78464 Konstanz, Germany.}
\affil[2]{Centre for the Advanced Study of Collective Behaviour, University of Konstanz, 78464 Konstanz, Germany.}
\affil[3]{Department of Biology, University of Konstanz, 78464 Konstanz, Germany.}

\begin{document}
	\maketitle

	\begin{abstract}
		
		Decades of scientific inquiry have sought to understand how evolution fosters cooperation, a concept seemingly at odds with the belief that evolution should produce rational, self-interested individuals. Most previous work has focused on the evolution of cooperation among boundedly rational individuals whose decisions are governed by behavioral rules that do not need to be rational. Here, using an evolutionary model, we study how altruism can evolve in a community of rational agents and promote cooperation. We show that in both well-mixed and structured populations, a population of objectively rational agents is readily invaded by mutant individuals who make rational decisions but evolve a distorted (i.e., subjective) perception of their payoffs. This promotes behavioral diversity and gives rise to the evolution of rational, other-regarding agents who naturally solve all the known strategic problems of two-person, two-strategy games by perceiving their games as pure coordination games. 
		
	\end{abstract}

	\section{Introduction}

Understanding the conditions that help to solve social dilemmas and promote cooperative behavior is of fundamental importance to understanding the biological and socio-economic world in which we live. It is also central to achieving solutions to our social and ecological problems, such as socio-economic \cite{Stiglitz2012}, and environmental inequality \cite{Hamann2018,Schell2020}, habitat destruction and biodiversity loss \cite{Cardinale2012,Bradshaw2009}, and the current environmental crisis \cite{Bak2021,Bodin2017,Gore2006}. Game theory, which considers strategic interactions among rational agents, has played a pivotal role in formulating the underlying problem for solving social dilemmas by introducing the notion of game equilibria \cite{Nash1950,Von1947}. In the Nash equilibrium of a game, no player can increase their payoff by changing strategy unilaterally \cite{Nash1950}. Tragedies of the commons, in which self-interest can result in the over-exploitation of public goods (such as a common resource), are an inevitable consequence of this notion of rationality when individuals’ and collective interests conflict, i.e., when a social dilemma exists \cite{Axelrod1981}. 

In contrast to this conclusion of classical game theory, cooperation is widespread in biological and social life and is a cornerstone of all forms of life, from multicellularity to social structure \cite{Axelrod1981}. An understanding of how, in contrast to the predictions of rationality in classical game theory, cooperation has evolved has been sought in evolutionary game theory \cite{Nowak2006,Szabo2007,Doebeli2005}. Research in this framework has typically abandoned the notion of rationality and has addressed how cooperation may evolve in the community of agents using simple behavioral rules. Consequently, the literature in evolutionary game theory has been shaped by the idea of bounded rationality \cite{Szabo2007}, according to which agents follow simple heuristics rather than rational reasoning to make decisions \cite{Simon1990}. Following this research program, theoretical work has successfully revealed diverse enforcement mechanisms that can solve or reduce tragedies of the commons \cite{Nowak2006,Szabo2007,Doebeli2005}, such as kin selection \cite{Hamilton1964}, direct \cite{Trivers1971,Hilbe2018,Schmid2021,Nowak1992} and indirect reciprocity \cite{Salahshour2022,Nowak2005,Ohtsuki2006a,Boyd1989,Schmid2021,Nowak1998,Milinski2002,Panchanathan2003,Sigmund2012,Sasaki2017}, network reciprocity \cite{Nowak1992,Ohtsuki2006b,Szabo2007}, punishment and reward \cite{Salahshour2019,Salahshour2021b,Szolnoki2015,Rand2009,Milinski2002,Szolnoki2014}, freedom of choice \cite{Salahshour2021a,Salahshour2021c}, or voluntary participation in the games \cite{Hauert2002}.

Here, we propose to approach the evolution of cooperation using a rather different perspective and by resorting to the notion of rationality. However, rather than taking the self-interested nature of rational decision-makers for granted, as rationality has been historically framed \cite{Hardin1968}, we study the evolution of subjective, other-regarding preferences in rational decision-makers. From behavioral economics and experimental studies, we know that humans and other animals often deviate from the classical notion of purely self-interested rational agents \cite{Fischbacher2001,Kocher2008,Herrmann2009,Fehr2002,Fehr2004,Fehr2005,Gintis2000,Thielmann2020}. Individuals appear to be influenced by other-regarding preferences and have subjective utilities that incorporate the welfare of others \cite{Fehr1999,Guth1998,Akcay2009}. This observation is captured by models developed within the so-called \textit{indirect evolutionary approach}, where preferences (or subjective payoff assessments) rather than strategies are subject to evolutionary forces \cite{Guth1998,Heifetz2007,Dekel2007,Samuelson2001,Huck1999,Akcay2009}. Such work has shown that payoffs can be `distorted' by subjective perceptions, allowing altruistic and other-regarding preferences to emerge as evolutionarily stable traits under certain conditions \cite{Heifetz2007,Huck1999,Guth1992,Dekel2007,Alger2019,Newton2018,Konigstein2000}. For instance, it has been shown how subjective preferences can explain a disposition towards altruistic behavior, such as fairness in the trust or ultimatum game \cite{Guth1998,Huck1999}, or reciprocal behavior \cite{Guth1992}. Other works in this framework have shown how the evolution of preferences can stabilize cooperation \cite{Dekel2007}. Furthermore, it has been argued that such a deviation from objective rationality can be generic, and thus a distorted perception of payoffs can often be advantageous when players have access to perfect information about others' types \cite{Heifetz2007}. 

We build on these insights and consider a model in which individuals have an evolvable other-regarding component to their subjective payoff. To do so, we assume that individuals exhibit the capacity to evolve a subjective (i.e., distorted) assessment of payoffs via an evolvable altruistic trait ($\delta$, from $0$ to $1$; representing a gradient from purely selfish to purely altruistic behavior), which they utilize to play the game. The altruistic trait, $\delta$, thus determines whether an agent exhibits other-regarding preferences \cite{Cooper2016,Fehr2002b,Bo2010} and will allow us to study the evolution of such preferences. Each individual is fully rational with respect to their own subjective payoffs, always playing the Nash equilibrium strategy profile that maximizes their subjective utility. However, their evolutionary success (reproductive fitness) depends only on the original (objective) payoff matrix of the game.  However, importantly, the agents' actual payoff is determined based on the original, objective payoff matrix of the game. 

By doing so, our framework emphasizes the often overlooked fact in evolutionary game theory that evolution often directly does not act on behavior. Rather it selects for the decision-making mechanism, such as emotional or perceptual factors, that shape behavior \cite{Budaev2019,Alger2019}. Our setup thus allows us to study the coevolution of behavior and subjective preferences: individuals evolve not in their observed behavior directly, but in how they subjectively perceive payoffs. For animals, like humans, this subjectivity may correspond to psychological factors, such as caring about others; but this feature could exist in much simpler organisms via any evolvable mechanism that modulates the relationship between the perception of, and actual, payoffs. 

Intuitively, one may expect that distorting the perception of reality in this way (i.e., having an altruistic trait above $0$) would undermine individuals' fitness in an evolutionary game, and thus, individuals should remain selfish. In the regime where the Nash equilibrium of the game is mutual defection, a tragedy seems inevitable, and thus, cooperation should not evolve. However, in contrast to such rationale, previous work has provided existence theorems according to which such deformations of payoff can be generically beneficial in the sense that there exist some deformations of payoffs that can increase an individual's fitness \cite{Heifetz2007}. We demonstrate that such altruistic deformations of payoff can indeed give rise to a diversity of rational personalities, and this diversity can serve as a way to the evolution of a broad spectrum of cooperative behavior in all the possible two-person, two-strategy games. 

Notably, the assumption that individuals can recognize others' preferences (or altruism) is crucial. It is akin to having type-recognition via phenotypic cues, language, or reputation \cite{Bowles2004,Henrich2007,Gintis2000b,Jansen2006,Traulsen2007,Salahshour2020,Gardner2010,Mailath2006}. This makes our model intimately connected to models of indirect reciprocity in the usual framework of evolutionary game theory. However, instead of studying indirect reciprocity based on rigid behavioral rules, our framework allows us to study whether conditional strategies similar to those observed in models of indirect reciprocity can evolve in rational agents and in a self-organized way.

We begin by showing that subjectively rational individuals show behavioral diversity, and mathematically identify four basic strategic `personality types', concerning altruistic incentives, as all the possible types in two-person two-strategy games. The usual notion of rationality in game theory (self-interested rationality) corresponds to one of these personality types, which can be considered objective rationality. Following this, using a simple evolutionary model, we show that objective rationality is evolutionarily unstable, and that individuals evolve into multiple coexisting personality types in both well-mixed and structured populations. Importantly, in our framework, selection does not directly act on strategies but rather on the decision-making mechanism (controlled by $\delta$) which can be applied to any game structure. We show that subjectively rational individuals not only solve social dilemmas but also solve all the strategic problems contained in two-person-two-strategy games by transforming them into pure coordination games, in the sense that the subjective game that the individuals play is always a coordination game.

Like many other mechanisms, cooperation is dependent on the evolution of conditional strategies. Unlike many other mechanisms, here cooperation emerges as a purely rational solution to the subjectively perceived games, turning them into effective coordination problems due to the evolved distorted perception of the games. This important point is missing in other models of indirect reciprocity which do not consider rational decision-making. The reciprocity underlying the evolution of cooperation in our framework is not hard-coded in agents' behavior (strategies). Rather, it evolves naturally due to the fact that, under perfect information, one can change the equilibrium behavior of an opponent by changing their own valuation of the game payoffs. This leads to the evolution of \textit{rational} conditional cooperators who selectively cooperate with those who value their game in an altruistic way. While core to many previous mechanisms for the evolution of cooperation (such as direct \cite{Trivers1971,Hilbe2018,Schmid2021} and indirect reciprocity \cite{Salahshour2022,Nowak2005,Ohtsuki2006a,Boyd1989,Schmid2021,Nowak1998,Milinski2002}), unlike others, the conditional cooperation in our model is not hard-coded in behavior. Rather, it evolves in a self-organized way even in a well-mixed population and in the absence of any other cooperation-favoring mechanism. We call this mechanism rational reciprocity which can provide an avenue to explore the notion of indirect reciprocity.

Moreover, we compare the evolution of cooperation among rational individuals with one of the most studied mechanisms for the evolution of cooperation: network reciprocity \cite{Szabo2007,Nowak1992,Ohtsuki2006b}, which can promote cooperation in a structured population of boundedly rational individuals (i.e., agents following rigid behavioral rules \cite{Szabo2007}). We show that cooperation among subjectively rational individuals, even in a well-mixed population, is more robust than the latter, in the sense that it evolves under broader conditions and more diverse game theoretic scenarios. Furthermore, the fact that the evolution of cooperation in subjective rationality results from rational decision-making (as individuals always play the Nash equilibrium of their subjective game) leads to the notable result that a simple replicator dynamic can explain the dynamics of the system, even in a structured population. This starkly contrasts the situation in bounded rationality, where the effect of population structure is vital, and even an approximate analytical description requires sophisticated analytical methods \cite{Szabo2002,Ohtsuki2006a}.

Our approach can thus be seen as a general theoretical framework that integrates ideas from evolutionary game theory, the indirect evolutionary approach, preference evolution, and rationality-based game theory, providing a plausible route for the evolution of universal forms of cooperation in some contexts.

\section{The Model and Setup}

In our model, we consider a population of agents playing a symmetric two-person two-strategy game (with strategies, $s\in{C,D}$, corresponding to `cooperation' and `defection', respectively). Each individual's subjective payoff depends on their evolved altruistic trait, $\delta_i\in[0,1]$, such that its subjective payoff is the weighted sum of its payoff and its opponent's payoff; $U^s_i(s_i,s_j)=(1-\delta_i)U(s_i,s_j)+\delta_iU(s_j,s_i)$, where $U$ is the objective payoff function of the game, and $s_i$ and $s_j$ are the strategies of, individual $i$ and individual $j$, respectively. Agents are rational and always play the Nash equilibrium of the game based on their subjective payoff function. The usual notion of rationality in game theory, which we refer to as selfish rationality, occurs when $\delta_i=0$. We consider one-shot games in both a well-mixed population (in which the evolution of cooperation is most restrictive) and a structured population, and assume individuals have perfect information about others' altruism. This assumption can hold in diverse contexts, such as small groups, structured populations, or when a reputation mechanism exists \cite{Bowles2004,Henrich2007,Gintis2000b} In contrast to the usual usage of bounded rationality, in our framework, agents are not defined by their strategies (e.g., cooperation and defection in the Prisoner's Dilemma). Rather, the agent's decision-making mechanism, controlled by a parameter $\delta$, is subject to evolution (in keeping with the indirect evolutionary approach \cite{Guth1998}). This fact allows us to study the same decision-making mechanism in different environments. Thus, we do not limit ourselves to the Prisoner's Dilemma and instead study three games, each belonging to one of the classes of two-person two-strategy games: the Prisoner's Dilemma, which belongs to the pure dominance class; the Snowdrift game, which belongs to the anti-coordination class; and the Stag Hunt Game, which belongs to the coordination class. These games are defined by the objective payoffs, the reward for mutual cooperation ($R$), the punishment for mutual defection ($P$), the temptation to defect ($T$), obtained when an individual defects while its opponent cooperates, and the sucker's payoff ($S$), obtained when an individual cooperates while its opponent defects (see Fig. \ref{Fig1}\textbf{a}). Individuals reproduce with a probability proportional to their objective payoff. Offspring inherit the altruistic trait $\delta$ of their parent, subject to mutations; with mutation rate $\nu$, offspring are created with $\delta\in[0,1]$ chosen uniformly at random (although, it should be noted that, in contrast to some previous models \cite{Hilbe2013}, our results are robust with respect to how mutations are introduced; see Methods for further details of the model). We study the evolutionary model using both agent-based simulations and replicator dynamics, as detailed in the Methods Section.

We note that several points differentiate between our approach and the usual approach in evolutionary game theory. Since such points are not explicit model assumptions, but rather are driven by our ``rationality-'' and subjectivity-based'' setup we explicitly discuss them. Firstly, while the objective game is the same for all, different agents perceive the game differently, and subjectively play different games. Moreover, since the altruistic trait of an individual affects its valuation of the game, the same individual essentially plays different subjective games when playing with different opponents (because the subjective payoff values of the opponent depend on their altruistic trait). Besides, individuals can evolve so that their subjective payoff depends on the opponent's payoff. Thus, while we limit the objective game to be a symmetric game, the subjective game can be, and often is, asymmetric. Since our agents are rational, and play the Nash equilibrium of their games, we do not need to take such complexities into account, for instance, by prescribing behavioral rules for agents to make decisions in each such possible game. In the following, we start with a formal analysis to drive possible rational personalities and then appeal to evolutionary models to study evolutionary outcomes.

\section{Results}

\subsection{Classification of personality types and games}

We begin by asking whether the usual notion of rationality is evolutionarily stable. To address this question, we consider a population of selfish rational individuals (i.e., $\delta_i=0$ for all the individuals, $i$) who play a Prisoner's Dilemma game (specified by the condition $S<P<R<T$) in a well-mixed population. A simple argument may provide two predictions. First, ``caring" about others (i.e., $0<\delta_i$) undermines individuals' fitness. Thus, individuals' altruism should not increase, and they should remain selfish rationals. Second, given that the Nash equilibrium of the Prisoner's Dilemma is mutual defection, a tragedy should be inevitable for $R<T=5$, and no cooperation should evolve in this regime. However, neither of these conclusions is correct.

The evolution of the average value of the altruistic trait in the population, $\bar{\delta}$ (altruism for short), and the fraction of agents exhibiting the cooperative strategy $\rho_C$, for three different values of the benefit of cooperation, $R$, over time is shown in Fig \ref{Fig2}\textbf{a} (a similar behavior is exhibited via replicator dynamics; see Supplementary Note 1). Although the Nash equilibrium is defection for $R<5$, we find that cooperation nonetheless evolves for $R>4$. Why is this the case? The first clue to the solution of this apparent paradox is provided by the histogram of the strategies played by and against individuals with a given altruism in Figs. \ref{Fig2}\textbf{c} and \ref{Fig2}\textbf{d}. As expected intuitively, individuals with high $\delta$ cooperate (Fig. \ref{Fig2}\textbf{c}). However, counter-intuitively, high $\delta$ also leads to receiving more cooperation, while individuals with low $\delta$ tend to experience defection (Figs. \ref{Fig2}\textbf{d}). This suggests that the evolution of cooperation can emerge via the self-organized punishment of selfishness via withholding cooperation.

\begin{figure*}
	\centering
	\includegraphics[width=1\linewidth, trim = 2 0 1 0, clip,]{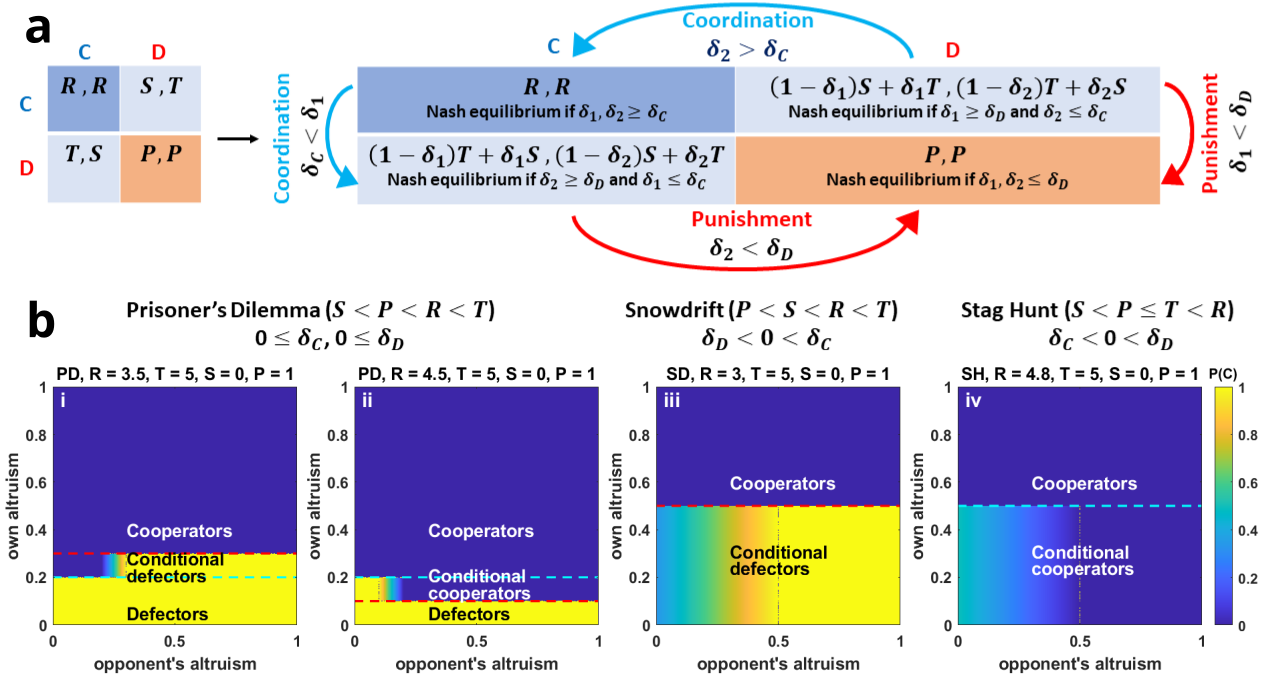}
	\caption{Subjective rationality in symmetric two-person two-strategy games. \textbf{a}: A symmetric two-person, two-strategy game is defined by two strategies ($C$ and $D$) and four payoff parameters. The first letter in each cell shows the payoff of the row player, and the second letter shows the payoff of the column player. Caring for others transforms the game into a game where non-diagonal elements of the payoff matrix are a linear superposition of one's own and the opponent's payoff. The coordination force induces cooperation when the value of the altruistic trait of the individual is larger than $\delta_C=\frac{T-R}{T-S}$, and the punishment force induces defection when it is smaller than $\delta_D=\frac{P-S}{T-S}$. Players' personality types can be classified based on their coordination and punishment force. \textbf{b}: The probability of cooperation in the Nash equilibrium of the games as a function of own and opponent's altruistic trait value are color plotted. $\delta_C$ (red) and $\delta_D$ (blue) are superimposed. Prisoner's Dilemma shows two phases for $\delta_D<\delta_C$ and $\delta_C<\delta_D$. In the former (i) conditional defectors who punish altruists and reward selfishness coexist with (unconditional) cooperators and defectors. In the latter (ii), conditional cooperators who reward altruists and punish selfishness replace conditional defectors. Cooperation evolves only in this phase. In the Snowdrift game (iii), cooperators and conditional defectors, who defect with altruists and play a mixed strategy among themselves, coexist. In the Stag Hunt game (iv), cooperators and conditional cooperators, who cooperate with altruists but play a mixed strategy among themselves, coexist.
	}
	\label{Fig1}
\end{figure*}

How does such selective withholding of cooperation evolve in the first place? The mechanism behind this phenomenon can be considered rational reciprocity which can promote cooperative \cite{Guth1998,Dekel2007,Huck1999} or reciprocal \cite{Guth1992} behavior, resulting in the evolutionary instability of objective rationality \cite{Heifetz2007}. This mechanism is illustrated in Fig. \ref{Fig1}\textbf{a} and \ref{Fig1}\textbf{b}. An individual, $i$, who cooperates while its opponent, $j$, defects does not change its strategy if and only if $U_i^s(D,D)<U_i^s(C,D)$, which gives, $\delta_i<\delta_D\equiv\frac{P-S}{T-S}$. Thus, individuals with the value of altruistic trait higher than $\delta_D$ never punish defection. Similarly, the opponent changes to cooperation if and only if $U_j^s(C,C)>U_j^s(D,C)$, which gives, $\delta_j>\delta_C\equiv\frac{T-R}{T-S}$. As illustrated in Fig. \ref{Fig1}\textbf{a}, by comparing the value of altruistic trait of the players with $\delta_C$ and $\delta_D$, it is possible to infer the equilibrium of the game. 

The comparison of the altruistic trait of individuals with $\delta_C$ and $\delta_D$ allows us to sort individuals into four basic strategic ``personalities" concerning altruistic incentives;\\
$\bullet$ Unconditional cooperators - satisfy $\delta_C<\delta_i$ and $\delta_D<\delta_i$. They are fully altruistic and always cooperate.\\
$\bullet$ Unconditional defectors - satisfy $\delta_i<\delta_C$ and $\delta_i<\delta_D$. They are fully selfish and always defect.\\
$\bullet$ Conditional cooperators - satisfy $\delta_i>\delta_C$, so do not deviate from mutual cooperation provided their opponent cooperates, and $\delta_i<\delta_D$, so do not deviate from mutual defection provided their opponent defects. Such individuals, therefore, selectively cooperate with altruists and punish selfishness by defection. This type of strategic personality has been frequently reported in economic experiments \cite{Fehr2002,Thielmann2020,Fehr2005,Fischbacher2001,Kocher2008,Herrmann2009,Fehr2004} and motivated some to introduce the notion of strong reciprocity and it has been argued to play an essential role in human cooperation \cite{Gintis2000,Fehr2004b}.\\
$\bullet$ Conditional defectors - satisfy $\delta_D<\delta_i<\delta_C$. Conditional defectors deviate from both mutual cooperation and mutual defection, and their Nash equilibrium is the anti-coordination cooperation-defection pair. Conditional defectors are anti-social individuals who cooperate with the selfish and defect with altruists. This strategic personality type is reminiscent of anti-social punishers who selectively punish cooperators and have been observed in economic experiments \cite{Herrmann2008,Salahshour2022a}.

The above argument enumerates all the possible strategic personalities in two-person two-strategy games played by agents who employ subjective assessment of payoffs. In simplified cases where either $\delta_C$ (coordination class) or $\delta_D$ (anti-coordination class) become negative, strategic personalities simplify to two types: conditional defectors, $(\delta_C<)\delta<\delta_D$, and cooperators, $\delta>\delta_C(>\delta_D)$ in the anti-coordination class, and conditional cooperators $(\delta_C<)\delta<\delta_D$ and cooperators, $\delta>\delta_D(>\delta_C)$ in the coordination class (see Table \ref{tab:classes}). 

Importantly, these strategic personalities are not model assumptions (in contrast to the usual practice in evolutionary game theory models \cite{Szabo2007}). Our only model assumption is the rationality of the players, i.e., they always play the Nash equilibrium of their subjective game. Rather these personalities result from a mathematical analysis as a categorization of different personality types of rational agents in symmetric two-person two-strategy games.

Notably, the standard classification of two-person two-strategy games into dominance, anti-coordination, and coordination classes \cite{Szabo2007} maps onto the signs of $\delta_C$ and $\delta_D$ (see Methods for details):

- In the \textbf{pure dominance} class (e.g., Prisoner's Dilemma), both $\delta_C>0$ and $\delta_D>0$ hold, allowing all four personality types to exist.\\
- In the \textbf{anti-coordination} class (e.g., Snowdrift), one finds $\delta_D<0<\delta_C$, permitting only two personality types (cooperators and conditional defectors).\\
- In the \textbf{coordination} class (e.g., Stag Hunt), one has $\delta_C<0<\delta_D$, leading to coexisting cooperators and conditional cooperators.

A summary of these classes and personality types is provided in Table \ref{tab:classes}, clarifying the variety of strategic phenotypes in subjectively rational populations.

\begin{table*}[ht!]
	\caption{Classification of game types and associated personality types as a function of $\delta_C$ and $\delta_D$.}
		\begin{tabular}{c c c c}
			Game class & Conditions & Personality types & Example \\
			\hline
			Pure dominance & $\delta_C>0,\delta_D>0$ & UC, UD, CC, CD & Prisoner's Dilemma \\
			Anti-coordination & $\delta_D<0<\delta_C$ & UC, CD & Snowdrift \\
			Coordination & $\delta_C<0<\delta_D$ & UC, CC & Stag Hunt \\
		\end{tabular}
	\label{tab:classes}
\end{table*}

\subsection{Prisoner's Dilemma: Evolving cooperation through conditional cooperation}
The decomposition of the population into different types already reveals different evolutionary phases in the Prisoner's Dilemma (or, more generally, in the pure dominance class). For $\delta_D<\delta_C$, which gives $T+S>R+P$ ($R<4$ for the standard parametrization of the Prisoner's Dilemma used here), mathematically, only three types may exist: cooperators, defectors, and conditional defectors. The Nash equilibrium of the game for different levels of altruism of the two players in this regime is illustrated in Fig. \ref{Fig1}\textbf{b}(i). However, despite the mathematical possibility of the evolution of conditional defectors, they do not evolve because defectors receive a higher payoff than others and dominate the population.

The Nash equilibrium of the game for $\delta_D>\delta_C$ ($R>4$) is presented in Fig. \ref{Fig1}\textbf{b}(ii). In this regime, defectors, cooperators, and conditional cooperators may exist. The key to the evolution of cooperation is the existence of conditional cooperators who punish defection.

In this regime, these three strategies are found in the population and exhibit rock-paper-scissors-like cyclic evolutionary dynamics. This can be seen in Fig. \ref{Fig3}\textbf{a}, where we present the density plot of individuals' altruistic traits as a function of time (the replicator dynamics show a similar phenomenology. See Supplementary Note 1). Such a dynamic has been frequently observed in evolutionary models \cite{McNamara2010} and results when no can dominate in the population, nor can few types coexist in a mutualistic polymorphism. Rather, different types coexist in an antagonistic polymorphism. In our model, such oscillation occurs due to density-dependent changes in the payoffs of individuals with different altruism (cooperators, conditional cooperators, and defectors), as explained below.

In the absence of defectors, conditional cooperators occasionally withhold cooperation from each other but cooperate with cooperators. Consequently, cooperators receive a higher payoff due to not paying the cost of punishment, and increase in number. This reduces the fraction of conditional cooperators, makes the population vulnerable to defectors, and leads to a burst of defectors. Once defectors increase, conditional cooperators perform better due to selectively cooperating with cooperators and defecting with defectors and increase in frequency, and so on. We note that, as $R$ increases, and thus the strength of social dilemma decreases, the advantage of defectors over unconditional cooperators decreases. Thus, the frequency of unconditional cooperators increases in the population, leading to increasing average $\delta$ as a function of $R$ (Fig. \ref{Fig3}\textbf{e}).

This analysis indicates that conditional cooperation plays a pivotal role in the evolution of cooperation among rational individuals, and is consistent with the diverse forms of conditional cooperation that have been argued to underlie cooperative behavior among rational \cite{Guth1992,Guth1998} or boundedly-rational agents, such as in direct \cite{Hilbe2017} and indirect reciprocity \cite{Schmid2021,Riolo2001}, or voluntary contribution to public goods under perfect information \cite{Battu2020,Guttman2013}.

To evaluate how population structure affects the evolution of altruism and cooperation, in Fig. \ref{Fig3}\textbf{b}, we present the density plot of individuals' altruism as a function of time in a structured population. Weakened cyclic dynamics and a lower density of defectors are two key differences when compared with a well-mixed population. In Fig. \ref{Fig3}\textbf{c} and Fig. \ref{Fig3}\textbf{d}, we present the density plot of the time-average altruism of the individuals in, respectively, a well-mixed and structured population as a function of $R$. The population-average altruism and the proportion of cooperation as a function of $R$ are presented in Fig. \ref{Fig3}\textbf{e} and Fig. \ref{Fig3}\textbf{f}, respectively. Here, the results of the replicator dynamics are also shown which are in strong agreement with simulation results in finite populations for both well-mixed and structured populations. While a slightly higher proportion of cooperative behavior and slightly higher values of the altruistic trait are observed in a structured population in the regime where cooperation can evolve, the effect of population structure is not appreciable, and the replicator dynamics are in good agreement with both structured and well-mixed populations. As we will shortly see, the independence of evolutionary dynamics from population structure in a community of subjectively rational agents also holds in other game structures and starkly contrasts the situation in boundedly rational agents. This observation arises from the fact that the evolution of cooperation in the framework of subjective rationality results from rational decision-making (in the form of a diversified form of rationality), which, as intuitively expected, should be independent of population structure.

\subsection{Snowdrift game: Cooperation through marginal altruism}
So far, we have focused on the Prisoner's Dilemma as a prototypical example of social dilemmas. In the following, we illustrate how agents with an evolvable subjective perception of payoff matrices can trivially solve strategic problems belonging to the anti-coordination and coordination class and take over a population of selfish rational individuals.

We begin with the Snowdrift game (SD, also known as the Hawk-Dove game or the Chicken game), which belongs to the anti-coordination class. This game satisfies the condition $T>R>S>P$, and is widely used to study conflict among individuals \cite{Doebeli2005,Szabo2007}. The Snowdrift game also creates a social dilemma (when $2R>T+S$), as individuals are better off when everybody cooperates, but each individual can increase their payoff by defection. However, this is a softer social dilemma than that contained in the Prisoner's Dilemma as (a sub-optimal) cooperation is maintained in the Nash equilibrium \cite{Szabo2007,Doebeli2005}. The Nash equilibrium of the game for the standard payoff values of the game and different altruism of players is plotted in Fig. \ref{Fig1}\textbf{b}(iii). In the Snowdrift game, $\delta_D$ becomes negative. Thus, the condition $\delta_i>\delta_D$ is always satisfied, and individuals do not have an incentive for mutual defection and can be classified only based on $\delta_C$ into cooperators ($\delta_i>\delta_C$), who always cooperate, and conditional defectors ($\delta<\delta_C$) who defect if their opponent is altruistic, and otherwise play a mixed strategy.

The evolutionary dynamics of subjectively rational agents in the Snowdrift game are studied in Fig. \ref{Fig4}\textbf{a}, where the density plot of the value of altruistic trait of the individuals in a well-mixed population is shown (see Supplementary Note 1 for replicator dynamics, and Supplementary Note 2 for further details). The situation is similar in a structured population (see Supplementary Note 1). We find that evolution gives rise to a population of marginally unconditional cooperators at the boundary of unconditional cooperation defined by $\delta=\delta_C$. This is the case because lower $\delta$ can lead to payoff loss in confrontations, and higher $\delta$ can lead to consistent exploitation by the selfish. Clearly, the mechanism underlying the evolution of cooperation in the Snowdrift is different from that in the Prisoner's Dilemma, as it does not involve conditional cooperation.

\subsection{Stag Hunt game: Avoiding coordination failure}

We move on to the Stag Hunt game (SH) which belongs to the coordination class and is defined by the condition $R>T\geq P>S$. This game is used to study coordination dilemmas: while mutual cooperation is the socially optimal Nash equilibrium (i.e., leads to the highest total payoff), mutual defection is the risk-dominant Nash equilibrium, and a population of individuals playing such a game show bistability and may settle for the inferior option (mutual defection). For this game, when played among subjectively rational agents, we have $\delta_C<0$, and thus, agents can be classified to cooperators ($\delta_i>\delta_D$) and conditional cooperators ($\delta_i<\delta_D$). The Nash equilibrium of this game when played among subjectively rational agents is presented in Fig. \ref{Fig1}\textbf{b}(iv). Cooperators always cooperate. However, conditional cooperators cooperate with cooperators and play a mixed strategy among themselves. As mutual cooperation endows the highest individual payoff, evolution leads to a monomorphic population of cooperators in this game. This can be seen in Fig. \ref{Fig4}\textbf{b}, where the density plot of individuals' altruism in a well-mixed population is presented (see Supplementary Notes 1 and 2 for details). Here, we have fixed, $R=3$, $P=1$, and $S=0$, and varied $T$ from $1$ to $5$. At $T=3$, the structure of the game changes to a Prisoner's Dilemma. At this point, $\delta_C$, plotted by the dashed red line, starts to be positive. This leads to the emergence of conditional cooperators and defectors in the population. At $T=R+P-S=4$, cooperators cease to dominate the population.

\subsection{Comparison with bounded rationality and network reciprocity}

To summarize, in Fig. \ref{Fig5}, we compare the performance of subjectively rational agents with boundedly rational agents in the three different games. In a well-mixed population, a population of boundedly rational agents is expected to converge to the Nash equilibrium of the game (dashed black line). However, deviation from the Nash equilibrium occurs in a structured population due to network reciprocity, a well-known enforcement mechanism in a population of boundedly rational agents \cite{Szabo2007,Nowak1992,Ohtsuki2006a}. While it is broadly known that population structure can promote cooperation in the Prisoner's dilemma, this is only the case for certain evolutionary dynamics. For instance, while cooperation can evolve when individuals reproduce with a probability proportional to the exponential of their payoff (Structured I), it does not evolve when individuals reproduce with a probability proportional to their payoffs (Structured II). By contrast, robust cooperation evolves in a community of subjectively rational agents and for a broader range of parameter values. Furthermore, the positive impact of population structure on the evolution of cooperation in a community of boundedly rational agents depends on the strategic problem at hand. For instance, while population structure can be detrimental to the evolution of cooperation in the Snowdrift game in a community of boundedly rational agents \cite{Hauert2004}, it does not appreciably impact the evolution of cooperation, especially for $R$ notably larger than $1$ in a community of subjectively rational agents. Similarly, subjective rationality can outperform network reciprocity in the Stag Hunt game (see Supplementary Note 3 for more details).

In Fig. \ref{Fig5}, the results of replicator dynamics for subjectively rational individuals are also shown. Replicator dynamics are expected to be an exact solution only in an infinite, well-mixed population and not in a structured population \cite{Nowak2006b}. Nonetheless, a very high agreement between the results in a structured population and replicator dynamics is observed (simulation results in finite well-mixed populations are similar). Such a level of agreement is not achievable for boundedly rational agents, where sophisticated methods are required to approximate the evolution of cooperation in structured populations (e.g., \cite{Szabo2002,Ohtsuki2006a}). As noted above, this results from the fact that rational decision-making, which is expected to be universal and independent of population structure, underlies the evolution of cooperation in subjective rationality.

\section{Discussion}

We have shown that allowing subjective assessment of payoffs and evolvable other-regarding preferences provides a unifying explanation for the emergence of cooperation in two-person two-strategy games under perfect information. By incorporating a trait that distorts payoffs to include opponents' utilities, individuals evolve diverse strategic personality types, including those that selectively reward altruists and punish selfishness. This transforms traditional social dilemmas and other coordination problems into effective coordination games, in the sense that individuals evolve to perceive their game as a coordination problem, and thus, cooperation can thrive.

Our approach falls under the indirect evolutionary approach, where preferences rather than strategies evolve \cite{Guth1998,Dekel2007,Heifetz2007}. While previous works in this tradition demonstrated the potential for non-selfish preferences to be evolutionarily stable and help resolve certain social dilemmas \cite{Guth1992,Huck1999}, our results show that such preferences can generically lead to robust and universal cooperation across all classes of two-strategy games. The essential ingredient is type recognition: individuals know the other player's altruism $\delta$, enabling them to adjust behavior conditionally. Without repeated encounters or separate reputation scores, the evolving subjective viewpoint provides enough information for effectively reciprocal behavior. In this sense, our model is closely related to known mechanisms like direct or indirect reciprocity, or partner choice, but achieves them in a one-shot setting by evolving preferences that make cooperation a rational response.

In this regards, our work can be considered a new dimension to models of indirect reciprocity \cite{Salahshour2022,Nowak2005,Ohtsuki2006a,Boyd1989,Schmid2021,Nowak1998,Milinski2002}. While the indirect reciprocity literature has focused on rigid behavioral  \cite{Nowak2005,Nowak1998} or explicit moral assessment rules \cite{Panchanathan2003,Sigmund2012,Sasaki2017,Ohtsuki2006a} which are used to determine good and bad behavior, rational conditional cooperators in our model are not introduced by model assumptions, nor are they defined as rigid behavioral rules. Rather these strategies are found by a mathematical analysis and evolve in a self-organized way. The underlying mechanism behind the evolution of such rational cooperative strategies is the fact that rational agents can change the behavior of their opponents by changing their own subjective valuation of the game, and thus effectively playing different games with different opponents, which admit different sets of rational choices (behavior).

Empirical studies have repeatedly observed that humans deviate from self-interested rationality and instead show strong reciprocity, conditional cooperation, and other-regarding preferences \cite{Fehr2002,Gintis2000,Fehr2004b,Fischbacher2001,Herrmann2009,Kocher2008,Thielmann2020}. Our findings suggest a simple evolutionary explanation: subjective payoff assessment (caring about others) can be favored by natural selection since, under broad strategic scenrios, it can lead to better evolutionary outcomes. Hence, what appears as deviations from classical rationality can be understood as alternative stable equilibria of an indirect evolutionary process.

An implicit assumption of our work is mutual recognition of preferences, as in our analysis agents have access to perfect information about the altruism of their opponent. This assumption can be a valid assumption in certain empirical contexts, such as structured populations, small-scale societies, or when a reputation mechanism \cite{Bowles2004,Henrich2007,Gintis2000b} or symbolic or phenotypic cues are at work \cite{Gardner2010}. In other contexts, incomplete information and uncertainty about others' types and intentions may be common, potentially leading to deviations from the predicted equilibria. Future studies could address this limitation by incorporating models of incomplete information and exploring how uncertainty and learning dynamics affect the evolution of cooperation, behavioral diversity, and rationality under such limitations.

Finally, our model is deliberately simple, focusing on symmetric two-strategy games. Extending the approach to richer games, multiple strategies, and more complex population structures can offer further insights. Nevertheless, the central message remains: by allowing preferences themselves to evolve, we discover that rational agents need not be selfish. Subjective rationality—emergent from evolutionary pressures—provides a plausible route for the evolution of altruistic behavior and stable cooperation in a wide class of social dilemmas.

\section{Material and Methods}

\subsection*{The Model and Simulations}

We consider a population of rational individuals playing a symmetric two-person two-strategy game, with strategies, $s\in{C,D}$ and the payoff matrix, $U(s,s^*)$, where $s$ is the focal individual's strategy and $s^*$ is the strategy of its opponent. Each individual, $i$, has an altruism, $\delta_i$, such that its subjective payoff is the weighted sum of its payoff and its opponent's payoff, $U_i^s=(1-\delta_i)U+\delta_i U$. Individuals are rational and play the Nash equilibrium of the game based on their subjective payoffs. In the main text, we consider a case where individuals play the mixed strategy Nash equilibrium if the game has a mixed strategy Nash equilibrium. Otherwise, they play a randomly chosen pure strategy equilibrium. However, to take the one-shot nature of the interactions into account, when a mixed strategy Nash equilibrium exists, individuals gather payoff by playing one of their strategies with a probability proportional to the mixed strategy Nash equilibrium. This choice does not affect the results, as for large enough populations, the expected payoffs of individuals with altruism $\delta$ converge to the expected payoff of a mixed strategy Nash equilibrium. Thus, the results are similar to that when individuals' payoff is determined based on the average expected payoff of the mixed strategy Nash equilibrium.

At each time step, all the individuals play the game, gather payoffs, and reproduce with a probability proportional to a selection parameter $\beta=0.5$ multiplied by their payoffs. Importantly, the actual payoff of individuals is determined based on the original payoff matrix of the game, $U$, and not their subjective payoffs. Offspring inherit the altruistic trait of their parent subject to mutations. Mutations occur with probability $\nu$, in which case the altruistic trait of the offspring is set to a value $\in[0,1]$ chosen uniformly at random. All the simulations begin with selfish-rational individuals, i.e., $\delta_i=0$ for all $i$. Altruistically rational individuals evolve when the value of altruistic traits grows strong enough to affect the behavior of the individuals.

We consider both a well-mixed population and a structured population. In a well-mixed population, at each time step, individuals are paired at random to play the game, after which all the individuals' strategies are updated simultaneously. In the structured population, individuals reside on a lattice and play the game with all their neighbors on the network. After playing the game, individuals imitate the strategy of one of the individuals in their extended neighborhood with a probability proportional to the exponential a selection parameter $\beta=0.5$ (consistently with many studies) times its payoff, $\exp(\beta \pi)$) (this algorithm is called Structured I in Fig. \ref{Fig5}). The extended neighborhood of an individual is composed of the focal individuals together with their neighbors. For the population network, we consider a first nearest-neighbor square lattice with periodic boundaries and Von Neumann connectivity. In Fig. \blue{5}, we also consider an alternative evolutionary dynamics, denoted by Structured II, in which individuals reproduce with a probability proportional to their payoffs. In Supplementary Information (Supplementary Note 4), we further study the evolutionary dynamics where individuals reproduce with a probability proportional to their payoff and show that all the results remain valid in this case.

Apart from the game payoffs and selection parameter, $\beta$, the evolutionary model has only two parameters: mutation rate, $\nu$, and the population size, $N$. In all the simulations performed in the main text, we have set $\nu=10^{-3}$ and $N=10,000$. In the Supplementary Information, the effect of population size (Supplementary Note 5) and mutation rate (Supplementary Notes 6) is studied, and it is shown that smaller population size and mutation rates can lead to the fixation of one of the types due to random drift, which can give rise to intermittent dynamics between periods of high and low altruism and cooperation. Furthermore, while some past studies have suggested the outcome of evolutionary dynamics may depend on how mutations are introduced \cite{Hilbe2013}, in Supplementary Note 6, we show that our results are robust under different ways mutations are introduced into the model. 

Finally, in Supplementary Information (Supplementary Note 7) we study the effect of different notions of equilibrium selection \cite{Harsanyi1998}, where individuals play the risk-dominant or payoff-dominant Nash equilibrium, and show that they lead to similar results. The only difference that different notions of equilibrium selection can have occurs in the anti-coordination class. In this case, when individuals play a pure strategy Nash equilibrium (irrespective of whether the Nash equilibrium is payoff-dominant, risk-dominant, or none), the evolutionary dynamics give rise to a population of conditional defectors who anti-coordinate on heterogeneous cooperation-defection pairs and avoid paying the cost of anti-coordination failure.

\subsection{Boundedly rational agents in a structured population simulation}

In Fig. \ref{Fig5}, we have compared subjectively rational agents with one of the well-known mechanisms for the evolution of cooperation in boundedly rational agents: population structure. Here, all the details of the simulation are exactly the same for boundedly rational and subjectively rational agents. The only difference is that, while subjectively rational agents have a heritable trait $\delta$, based on which they calculate their subjective payoffs, boundedly rational agents have a hard-wired heritable strategy, cooperation, $C$, and defection, $D$. 

For clarity, we repeat other details of the evolutionary simulation, which is a standard highly replicated procedure \cite{Szabo2007}. Each individual resides on a first nearest-neighbor square lattice with periodic boundaries and Von Neumann connectivity and plays the game with all their neighbors. After playing the game, individuals imitate the strategy of one of the individuals in their extended neighborhood with a probability proportional to a selection parameter $\beta=0.5$ times its payoff. We have dubbed this algorithm, which is a standard algorithm \cite{Szabo2002}, structured I in Fig. \ref{Fig5}. We have also considered a second algorithm in which individuals reproduce with a probability proportional to their payoffs (structured II). The extended neighborhood of an individual is composed of the focal individuals together with their neighbors.

We note that, while our subjectively rational agents know the altruism of their opponent (an implicite assumption behind the assumption that they play the Nash equilibrium of their subjective game), such an assumption is not implemented for boundedly rational agents. The reason is that our aim is to compare our mechanisms with one of the most intensely studied and standard mechanisms for the evolution of cooperation and we have chosen a standard evolutionary algorithm that does not rely on any sort of recognition, but rather on spatial affinity of individuals. If for instance, we aimed to compare our model with models of indirect reciprocity, which rely on recognition mechanisms, then such an assumption would have been essential in both cases, and one could have argued our comparison was more fair.

\subsection*{The game structures}

We study three games, each belonging to one of the classes of two-person two-strategy games;\\
$\bullet$ Prisoner's Dilemma, which belongs to the pure dominance class \cite{Szabo2007}. This game satisfies $S<P<R<T$ and is widely used to study social dilemmas; Mutual cooperation provides the highest total payoff. However, each player can increase its payoff by defection. Consequently, the Nash equilibrium of the game is mutual defection which leaves all the players worse off than if they had cooperated. Such a ``tragedy of the commons'' is unavoidable when this game is played between purely rational agents. The base parameter values used are $T=5$, $S=0$, and $P=1$, and $R$ is varied between $1$ and $5$.\\
$\bullet$ Snowdrift game, which belongs to the anti-coordination class. This game is also known as the Hawk-dove or the Chicken game and belongs to the anti-coordination class \cite{Doebeli2005,Szabo2007}. This game satisfies $P<S<R<T$. Defection yields a higher payoff than cooperation if the opponent cooperates. Thence a temptation to defect exist. However, mutual defection leads to the least payoff. Consequently, the Nash equilibrium is a heterogeneous strategy pair where individuals anti-coordinate on defection and cooperation. In addition, this game admits a mixed strategy Nash equilibrium. Thus this game provides only a weak social dilemma since in its Nash equilibrium a sub-optimal cooperation is observed. This game is widely used in modeling conflict between individuals, such as a contest over a shared resource: individuals can escalate (defection) or conciliate (cooperation). A conflict is inevitable if both individuals escalate, leading to a pure cost to both individuals. Consequently, its in both individuals' best interests to avoid conflict. The base parameter values used for the Snowdrift game are $T=5$, $S=1$, and $P=0$, and $R$ is varied between $1$ and $5$.\\
$\bullet$ Stag Hunt game, which belongs to the coordination class. This game is widely used to model coordination problems and satisfies $S<P\leq T<R$ \cite{Szabo2007}. In the usual framing of the game, players can coordinate to hunt Stag (cooperation) or Hare (defection). Stag are more valuable. However, stag hunting requires more effort and is successful only if both players coordinate. However, players can receive a lower payoff from hunting hare irrespective of what their opponent does. This game has two pure strategy Nash equilibria, mutual cooperation provides the higher payoff and is the socially optimal outcome, and mutual defection provides the lower outcome but is the risk-dominant Nash equilibrium. In addition, this game admits a mixed strategy Nash equilibrium. The base parameter values used for the Stag Hunt game used here are $R=3$, $P=1$, $S=0$, and $T$ is varied between $1$ and $5$. We note that for $R>3$, this game becomes a Prisoner's Dilemma.

\subsection*{The relation with standard classes of two-person two-strategy games}

Symmetric two-person two-strategy games played between subjectively rational agents can be classified into three classes according to $\delta_C$ and $\delta_D$. Pure dominance class: $\delta_C\geq0$ and $\delta_D\geq0$, anti-coordination class, $\delta_D<0<\delta_C$, and coordination class, $\delta_C<0<\delta_D$. It is easy to see that the pure dominance class gives rise to four personality types. However, the two latter classes give rise to only two strategic personality types based on comparing the altruism of the individuals with either $\delta_C$ or $\delta_D$, illustrated in Fig. \blue{1}. The main difference between the coordination and anti-coordination classes is that there is no incentive to punish in the anti-coordination class. Thus mutual punishment can not be a Nash equilibrium. A final trivial case can be when $\delta_C\leq0$ and $\delta_D\leq0$. In this case, the only Nash equilibrium is mutual cooperation, independent of the altruism of the players. This can happen only if cooperation Nash dominates defection ($R>T$, and $S>P$). Given this game is trivial, we do not consider this case as an independent class.

The classification of the two-person two-strategy games introduced here largely coincides with the classification of the two-person two-strategy games played by selfish altruists, i.e., the standard classification of the symmetric two-person two-strategy games \cite{Szabo2007}. The only difference between the classification of the games played between altruistically rational agents and selfish rational agents is that the latter does not depend on the denominator, $T-S$, in the definition of $\delta_C$ and $\delta_D$. This classification can be reproduced by setting $\delta_1=\delta_2=0$, which requires replacing $\delta_C=\frac{T-R}{T-S}$ by $T-R$, and $\delta_D=\frac{P-S}{T-S}$ by $P-S$. This gives the pure dominance class satisfying, $T-R\geq0$ and $P-S\geq0$, where the only Nash equilibrium is mutual defection, anti-coordination class $P-S<0<T-R$, $C-D$, and $D-C$ together with a mixed strategy Nash equilibrium exist, and coordination class, $T-R<0<P-S$, where mutual cooperation and mutual defection, together with a mixed strategy are the Nash equilibria.

\subsection*{The mixed strategy Nash equilibrium}

A mixed strategy is a set of probabilities for playing different pure strategies. The support of a mixed strategy is the set of all the pure strategies which are played with nonzero probability. A mixed strategy Nash equilibrium is defined as a set of two mixed strategies, $(q_i, q_j)$ in which each strategy is the best response to the other strategy: $q_i = BR(q_j)$ and $q_j = BR(q_i)$. This condition is achieved if the payoff of all the strategies in the support of a mixed strategy is the same (indifference condition), and no other strategy outside of the support gives a higher payoff against the mixed strategy \cite{Salahshour2022}. In games with two strategies, the calculation of a mixed strategy is simplified to solving a set of linear equations to achieve the indifference of each player. To see this, assume player $i$ plays strategy $C$ with probability $q_i(C)$ and plays $D$ with probability $q_i(D)=1-q_i(C)$. The expected payoff of player $j$ from playing $C$ is $q_i(C)R^s_j+(1-q_i(C))S^s_j$ and its expected payoff from playing $D$ is $q_i(C)T^s_j+(1-q_i(C))P^s_j$. Setting these two equal gives for the mixed strategy Nash equilibrium of player $i$, $q^*_i(C)= \frac{P^s_j-S^s_j}{R^s_j+P^s-T^s_j-S^s_j}$. Here, a superscript $s$ refers to the subjective payoffs, $R^s_j=R$, $P^s_j=P$, $S^s_j=(1-\delta_j)S+\delta_jT$, and $T^s_j=(1-\delta_j)T+\delta_jS$. Similarly, the Nash equilibrium mixed strategy of player $j$ can be derived as $q^*_j(C)= \frac{P^s_i-S^s_i}{R^s_i+P^s-T^s_i-S^s_i}$. A curious and well-known property of mixed strategy Nash equilibrium is that a player's equilibrium strategy is determined by its opponent's payoffs. A similar observation also holds in pure strategy equilibrium and is central to rational reciprocity in games played between subjectively rational agents: by changing own values, one can change others' behavior.

\subsection*{Replicator-mutator dynamics}
The model in a well-mixed population can be solved in terms of the replicator-mutator dynamics \cite{Nowak2006b}. We begin by developing the replicator-mutator dynamics for continuous altruism, $\delta$, and then recover the equations for discretized altruism. The replicator-mutator equations govern the time evolution of the density of individuals with different values of altruistic trait, $\rho_t(\delta)$, and read as follows:
\begin{align}
	\rho_{t+1}(\delta)=\int_{0}^{1}W(\delta,\delta')\rho_t(\delta')\pi_t(\delta')/\bar{\pi_t}d\delta'.
\end{align}
Here, $\pi_t(\delta)$ is the expected payoff of individuals with altruism $\delta$ at time $t$, and $\bar{\pi_t}=\int_{\delta=0}^{\delta=1}\pi_t(\delta)\rho_t(\delta)d\delta$, is the average payoff of the population at time $t$. $W(\delta,\delta')$ is the mutation rate from an altruism $\delta'$ to altruism $\delta$. Under our assumption of uniform mutations, this can be written: $W(\delta,\delta')=\nu+(1-\nu)\Delta(\delta-\delta')$. Here, $\Delta(\delta-\delta')$ is the Dirac delta function (usually shown by $\delta$).

To use the replicator dynamics, we need to calculate the payoffs.
\begin{align}
	\pi_t(\delta)=\int_{0}^{1}\rho_t(\delta')\bm{q^*}(\delta,\delta')\bm{U}'\bm{q^{*t}}(\delta',\delta)d\delta'
\end{align}
Here, $\bm{U}'$ is the (objective) effective payoff matrix of the game. In the model considered in the main text, where individuals reproduce with a probability proportional to a selection parameter times their payoff, we have $\bm{U}'=\exp(\beta \bm{U})$ (where the exponential of the matrix $\bm{U}$ is defined as an element-wise exponential). When individuals reproduce with a probability proportional to their payoffs (considered in the Supplementary Information), we have $\bm{U}'=\bm{U}$. $\bm{q^*}(\delta,\delta')$ is the mixed (or pure) strategy Nash equilibrium of a subjectively rational agent with altruism $\delta$ when the agent plays a game with an agent with altruism $\delta'$. For two-person two-strategy games considered here, these can be represented by a vector, $\bm{q^*}=[q_C^*\quad q_D^*]$. A superindex $t$ indicates matrix transpose. The average payoff of the population can then simply be written as:
\begin{align}
	\bar{\pi_t}=\int_{0}^{1}\pi_t(\delta)\rho_t(\delta)d\delta.
\end{align}
This set of equations describes the dynamics of the system. To numerically solve these equations, we discretize the altruistic trait such that it can take $n$ possible values. This can be considered as an approximation of the integrals. However, such an approach is also of interest when the altruistic trait can take discrete values. The effect of discretization of the altruistic trait ($n$) is studied in the Supplementary Note. 1. The replicator dynamics for discretized altruistic trait can be written as:
\begin{align}
	\rho_{t+1}(\delta_k)=\sum_{k'=1}^{n}[\Delta_{k,k'}(1-\nu)+\nu/n]\rho_t(\delta_{k'})\pi_t(\delta_{k'})/\bar{\pi}.
\end{align}
Here, $k\in[1,n]$, where $n$ represents the number of possible values of the altruistic trait. A reasonable choice (which is also consistent with an approximation of the integral over continuous altruistic trait with a summation over discrete values of the altruistic trait), which we will use in numerical solutions, is linearly spaced values of altruistic traits in the interval $[0,1]$. $\Delta_{k,k'}$ shows the Kronecker delta function, which is equal to one if $\delta=\delta'$ and zero otherwise. The payoff of an individual with altruism $\delta_k$ reads as:
\begin{align}
	\pi_t(\delta_k)=\sum_{k'=1}^{n}\rho_t(\delta_{k'})\bm{q^*}(\delta_{k},\delta_{k'})\bm{U}\bm{q^{*t}}(\delta_{k'},\delta_{k}).
\end{align}
and the average payoff reads as:
\begin{align}
	\bar{\pi_t}=\sum_{k'=1}^{n}\pi_t(\delta_{k'})\rho_t(\delta_{k'}).
\end{align}

The numerical solutions of the replicator dynamics for $n=400$ are presented in \blue{3}\textbf{e} and \blue{3}\textbf{f} and Fig. \blue{5}, and show high agreement with simulation results. We note that the payoff accrued in the replicator dynamics is taken to be the expected payoff of the mixed strategy Nash equilibrium. However, in the simulations, to take the one-shot nature of interaction into account, we have chosen to consider a case where individuals play one of the two possible strategies, with a probability proportional to their probability in the mixed strategy Nash equilibrium. This assumption does not change anything other than increasing the finite size effects. This is the case because the expected payoff of individuals with altruism $\delta$, when averaged over all the individuals in the population, approaches that of the mixed strategy Nash equilibrium for large enough population sizes.

\section{Acknowledgment}

We are thankful to Arne Traulsen, Urs Fischbacher, and Raghavendra Gadagkar for insightful comments on an earlier version of the manuscript. The authors acknowledge funding from Deutsche Forschungsgemeinschaft (DFG, German Research Foundation) under Germany’s Excellence Strategy - EXC 2117-422037984, the Deutsche Forschungsgemeinschaft Gottfried Wilhelm Leibniz Prize 2022 584/22 (I.D.C.), the Max Planck Society, the European Union’s Horizon 110 2020 Research and Innovation Programme under the Marie Skłodowska-Curie Grant agreement no. 860949, the Struktur- und Innovations fonds für die Forschung of the State of Baden-Württemberg, the Path Finder European Innovation Council Work Programme no. 101098722, the Office of Naval Research Grant N0001419-1-2556.



\newpage

	\begin{figure}
	\centering
	\includegraphics[width=1\linewidth, trim = 10 0 20 10, clip,]{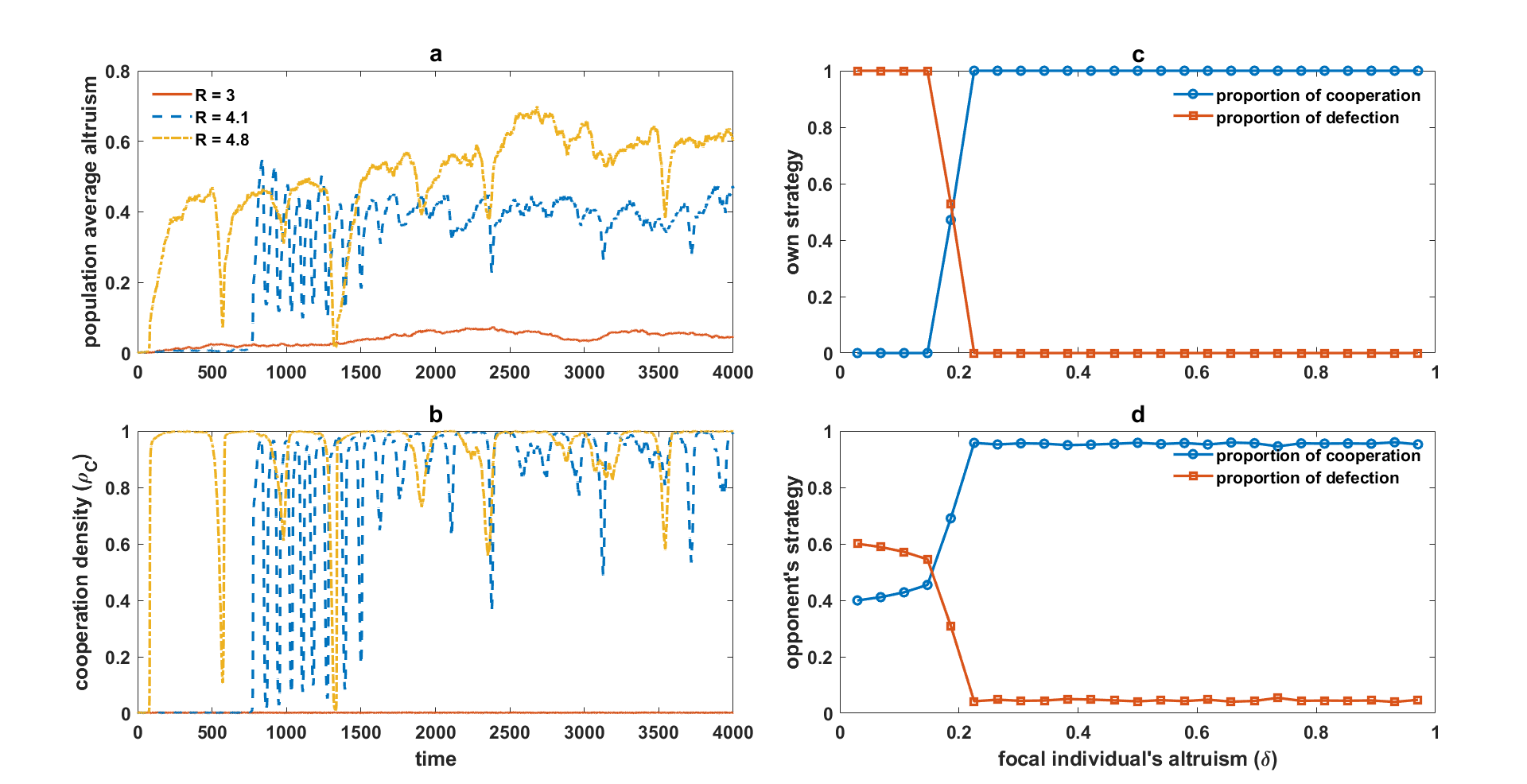}
	\caption{The evolution of altruism due to the rational punishment of selfishness. \textbf{a} and \textbf{b}: Population-average of individuals' value of the altruistic trait, $\bar{\delta}$ (\textbf{a}), and the density of cooperators, $\rho_C$ (\textbf{b}), as a function of time is plotted. The evolutionary simulation starts with a population of selfish rationals with zero value of the altruistic trait who fail to cooperate in the Prisoner's Dilemma. However, for $R>4$, altruist rationals evolve and dominate the population, and cooperation reaches a high level. Both the population average value of altruistic trait (altruism) and density of cooperation show fluctuations, suggesting a cyclic dominance of individuals with different altruism. \textbf{c} and \textbf{d}: The strategy played by (\textbf{a}) and against an individual (\textbf{d}) with a given value of the altruistic trait, $\delta$. Individuals with high values of altruistic traits cooperate, while those with low values of altruistic trait defect. Surprisingly, a higher value of altruistic traits also leads to receiving more cooperation and less defection. This curious phenomenon underlines the evolution of cooperation resulting from rational reciprocity. Parameter values: $N=10000$ and $\nu=10^{-3}$. A well-mixed population is used. Individuals play a Prisoner's Dilemma, $T=5$, $P=1$, $S=0$. In \textbf{a}, $R$ is indicated in the legend, and in \textbf{b}, $R=4.1$.}
	\label{Fig2}
\end{figure}

	\begin{figure}
	\centering
	\includegraphics[width=1\linewidth, trim = 60 0 470 2, clip,]{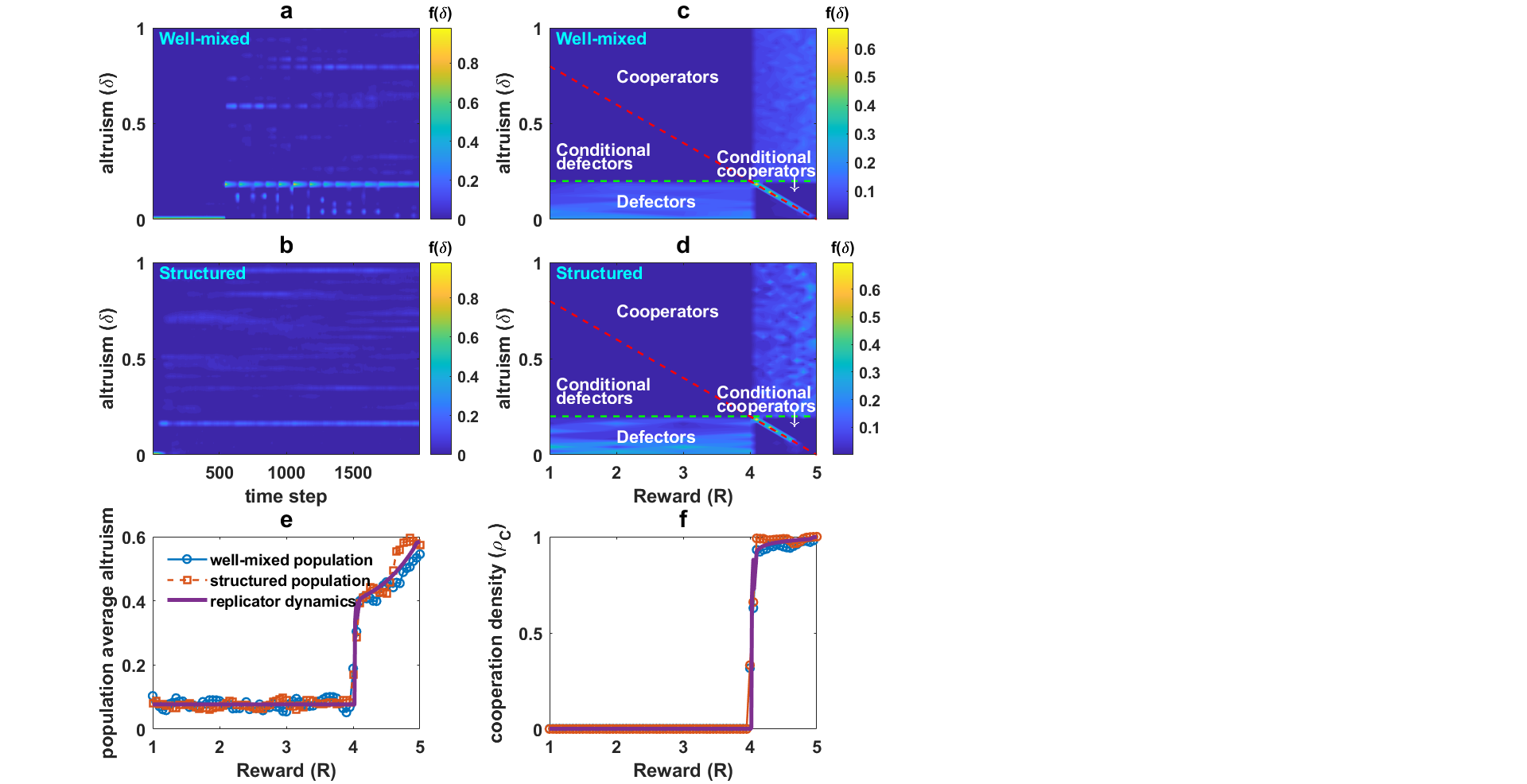}
	\caption{The evolution of altruistic rationality in the Prisoner's Dilemma. \textbf{a} and \textbf{b}: The density plot of the individuals' altruism as a function of time in a well-mixed (\textbf{a}) and structured (\textbf{b}) population. Individuals show behavioral diversity and can be decomposed into defectors with a low value of the altruistic trait, conditional cooperators, and cooperators with a high value of the altruistic trait, showing rock-paper-scissors-like dynamics. \textbf{c} and \textbf{d}: The density plot of the time-average value of the altruistic trait in a well-mixed (\textbf{c}) and structured (\textbf{d}) population as a function of the benefit of cooperation, $R$, are plotted. Comparison of individuals' value of altruistic trait with $\delta_C$ (green dashed) and $\delta_D$ (red dashed) allows identification of different personality types. Below, $R=T+S-P=4$, defectors dominate, and above $R=4$, cooperators, conditional cooperators, and defectors coexist. Increasing $R$ increases the density of cooperators and decreases the density of defectors and conditional cooperators, especially in a structured population. \textbf{e} and \textbf{f}: The population-average value of altruistic trait (\textbf{e}), and the proportion of cooperation, $\rho_C$ (\textbf{f}), in a structured and well-mixed population as a function of the benefit of cooperation, $R$ are plotted. The purple line shows the result of the replicator dynamics expected to be an exact solution of the model in a well-mixed population and in the infinite population limit. Parameter values: $N=10000$, $\nu=10^{-3}$. In (a), $R=4.1$. }
	\label{Fig3}
\end{figure}

	\begin{figure}
	\centering
	\includegraphics[width=1\linewidth, trim = 10 150 30 20, clip,]{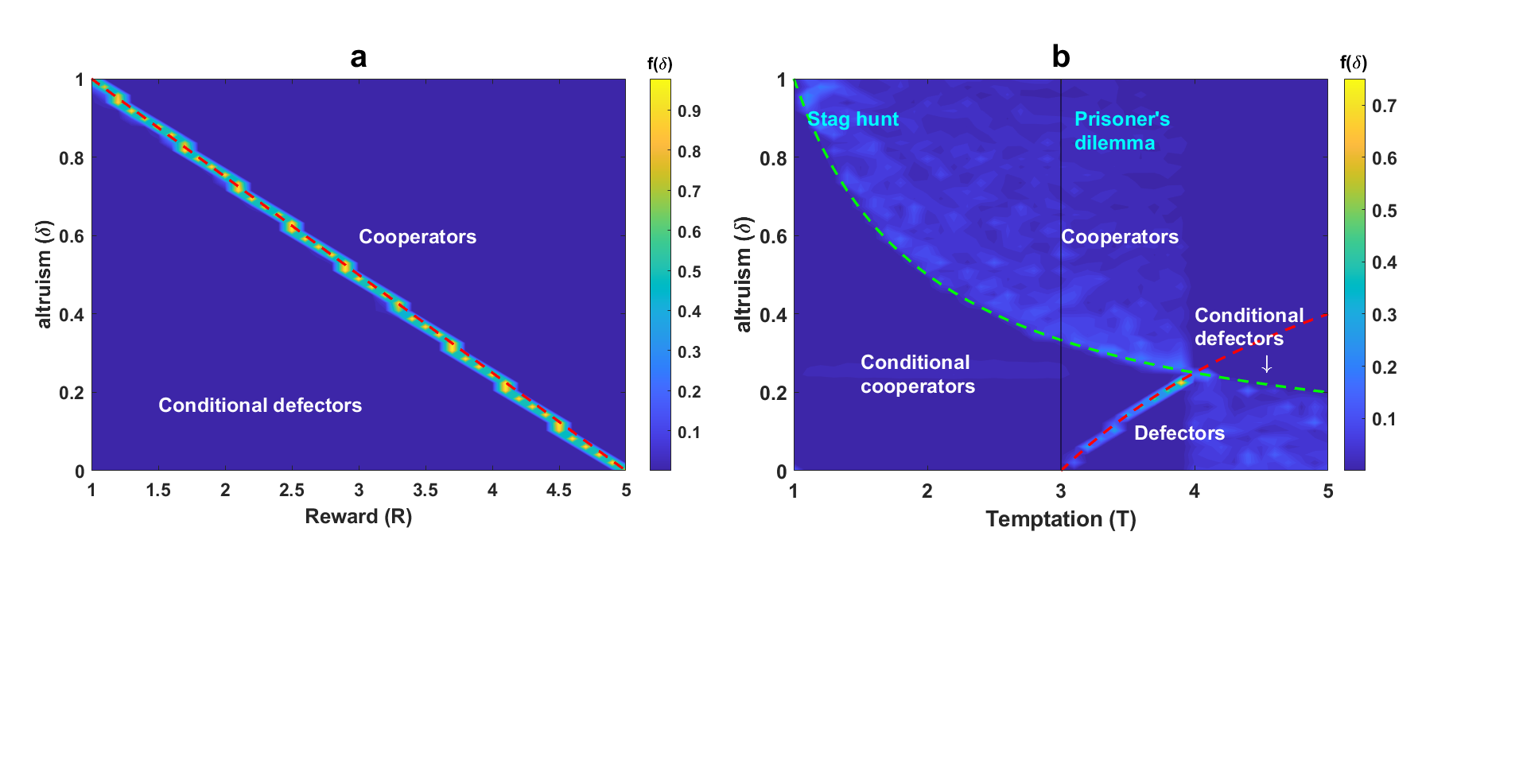}
	\caption{The evolution of cooperation by subjective rationality in the Snowdrift and Stag Hunt games. The density plot of the time-average value of the altruistic trait of the individuals as a function of the benefit of cooperation, $R$, in the Snowdrift game \textbf{a}, and the temptation to defect, $T$, in the Stag Hunt game \textbf{b}. \textbf{a}: In the Snowdrift game, individuals self-organize on the line defined by $\delta=\delta_D$ and are marginally unconditional cooperators. As a result, full cooperation evolves. \textbf{b}: In the Stag Hunt game, individuals evolve above the line defined by $\delta=\delta_C$. Consequently, a monomorphic population of unconditional cooperators evolves who successfully avoid coordination failure. Above $T=3$, the game becomes a Prisoner's Dilemma, and the coexistence of cooperators, unconditional cooperators, and defectors is observed. For too high $T$, defectors dominate. Parameter values: $N=10000$, $\nu=10^{-3}$.}
	\label{Fig4}
\end{figure}

	\begin{figure*}
	\centering
	\includegraphics[width=1\linewidth, trim = 80 280 60 5, clip,]{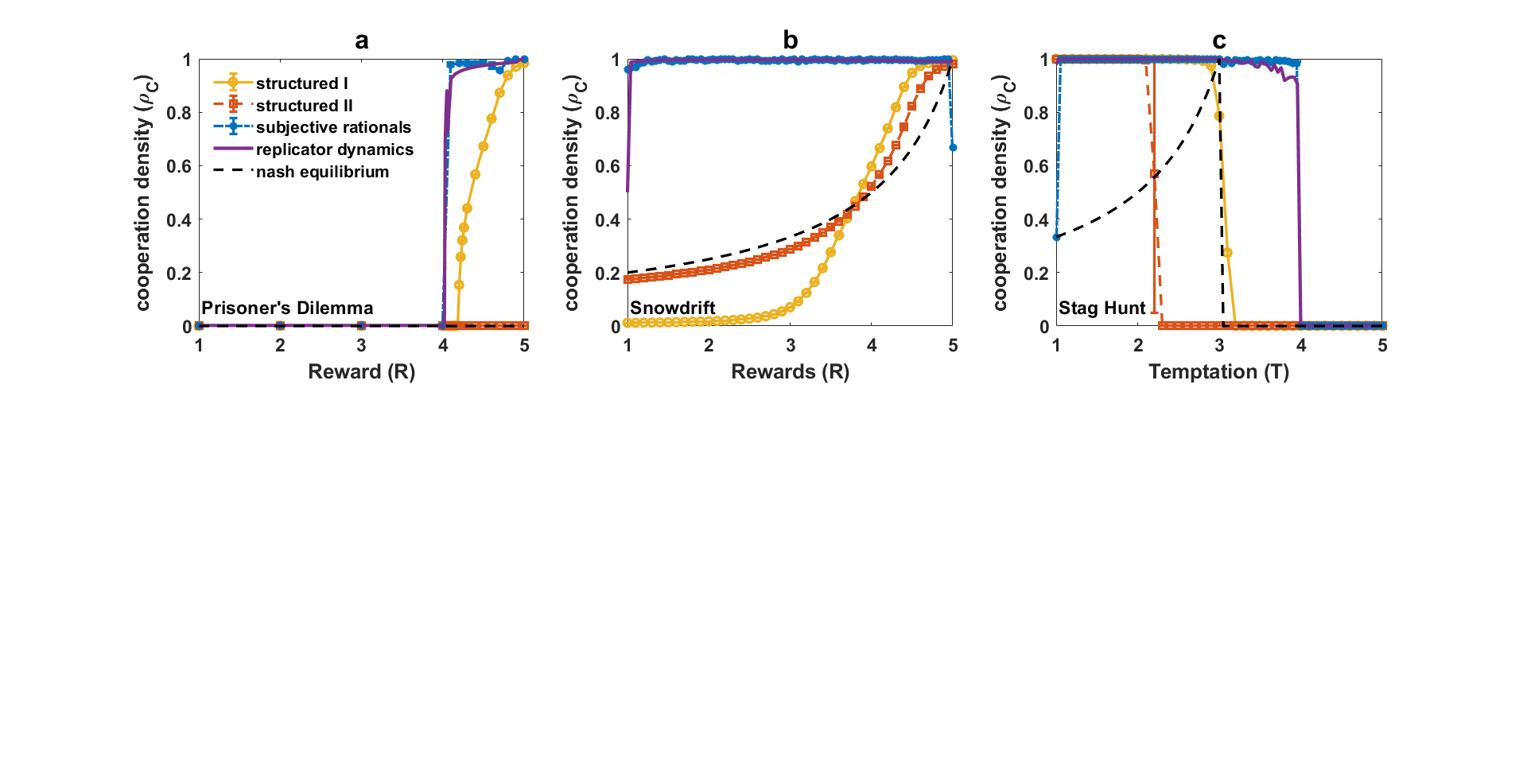}
	\caption{Comparison with network reciprocity. \textbf{a} to \textbf{c}: The outcome of evolutionary dynamics in a population of subjectively rational agents in a structured population (blue) and using replicator dynamics (purple, well-mixed population in the limit of large population size) is compared with that in a population of boundedly rational agents in a structured population (structured I, orange, and structured II, red) for Prisoner's dilemma (\textbf{a}), Snowdrift (\textbf{b}), and the Stag Hunt game (\textbf{c}). The Nash equilibrium of the game (mixed or pure), expected to occur in a well-mixed population, is plotted by a dashed black line. Structured I indicates an evolutionary algorithm in which individuals reproduce with a probability proportional to the exponential of a selection parameter $\beta=0.5$, times their payoff ($\exp(\beta \pi$)), and structured II indicates an evolutionary algorithm in which individuals reproduce with a probability proportional to their payoff (this algorithm is also used for subjectively rational individuals here). Subjectively rational agents outperform bounded rational agents and reach higher cooperation in all three games. Cooperation in a structured population remains close to that predicted by the replicator dynamics, indicating population structure does not play a significant role in promoting cooperation. Parameter values: $N=10000$ and $\nu=10^{-3}$. The population resides on a first-nearest-neighbor network with Von Neumann connectivity and periodic boundaries.}
	\label{Fig5}
\end{figure*}
	
\end{document}